**Serum 25-hydroxyvitamin D concentration is not associated with mental health among Aboriginal and Torres Strait Islander Peoples in Australia: a cross-sectional exploratory study**

Belinda Neo [1], Noel Nannup [2], Dale Tilbrook [3], Carol Michie [2], Cindy Prior [2], Eleanor Dunlop [4,5], Brad Farrant [2], Won Sun Chen [1], Carrington C.J. Shepherd [1,2], and Lucinda J. Black [4,5,*]

[1] Curtin Medical School, Curtin University, Bentley, Western Australia, Australia

[2] The Kids Research Institute Australia, Nedlands, Western Australia, Australia

[3] Maalinup Aboriginal Gallery, Caversham, Western Australia, Australia

[4] Curtin School of Population Health, Curtin University, Bentley, Western Australia, Australia

[5] Institute for Physical Activity and Nutrition (IPAN), School of Exercise and Nutrition Sciences, Deakin University, Geelong, Victoria, Australia

**Abstract**

**Objective:** To investigate the association between serum 25-hydroxyvitamin D [25(OH)D] concentration and mental health, measured using the Kessler Psychological Distress Scale 5 (K5), among Aboriginal and Torres Strait Islander Peoples.

**Methods:** We used cross-sectional data from the 2012-2013 Australian Aboriginal and Torres Strait Islander Health Survey. Multiple linear regression was used to test the association between serum 25(OH)D concentration and K5, adjusting for age, sex, education, remoteness, socioeconomic status, season of blood collection, smoking, and alcohol intake (n = 1,983). We also stratified the analysis by sex and by remoteness.

**Results:** There was no statistically significant association between serum 25(OH)D concentration and K5 in the total population, nor when stratified by sex. When stratified by remoteness, higher serum 25(OH)D concentration was statistically significantly associated with lower K5 scores among those living remotely (adjusted β: -0.18; 95% CI: -0.35, -0.01).

**Conclusions:** Serum 25(OH)D concentration was inversely associated with psychological distress only among those living remotely.

**Implications for Public Health:**

Given the prevalence of vitamin D deficiency and the observed association between serum 25(OH)D concentration and psychological distress among Aboriginal and Torres Strait Islander Peoples living remotely, public health strategies to improve vitamin D status among this population group are warranted.

**Keywords:** Vitamin D, mental health, Aboriginal and Torres Strait Islander Peoples, psychological distress

**Introduction**

Aboriginal and Torres Strait Islander Peoples have inhabited Australia for approximately 65,000 years.[1] The colonisation of Australia ~250 years ago resulted in the dispossession of land and forced removal of children and babies from Aboriginal and Torres Strait Islander families. The aftermath of colonisation has left a significant enduring impact on the mental health and wellbeing of Aboriginal and Torres Strait Islander Peoples.[2,3] Findings from the 2018-2019 National Aboriginal and Torres Strait Islander Health Survey showed that 17% of Aboriginal and Torres Strait Islander Peoples aged ≥ 2 years have anxiety, and 13% experienced depression/feelings of depression.[4] A higher proportion of females (35%) experienced high/very high levels of psychological distress compared to males (26%).[4]

Vitamin D deficiency is prevalent among Aboriginal and Torres Strait Islander Peoples, affecting 27% of adults aged ≥ 18 years.[5] The prevalence of vitamin D deficiency is higher among those living in remote areas (39%) compared to those in non-remote areas (23%).[5] Vitamin D may have a role in modulating mental health conditions as its receptors are present in brain regions involved in the pathophysiology of stress and mood disorders, such as depression and anxiety.[6-9] The postulated mechanisms of vitamin D on mental health conditions include the modulation of neurotrophic factors necessary for neuron viability and growth, the synthesis of neurotransmitters responsible for mood regulation, and the modulation of the immune system.[9-12] These mechanisms suggest that vitamin D status may be a modifiable mental health risk factor.

Previous findings from population-based observational studies examining the association between low serum 25-hydroxyvitamin D (25(OH)D) concentration and mental health conditions, such as depression, among various population groups, have been inconsistent:

some studies showed an inverse association,[13-15] but other studies showed no association.[16-18] The inconsistencies between those studies could be attributed to the use of an uncertified assay to test serum 25(OH)D concentration,[13, 15-17] the use of different tools to assess mental health conditions, and the inclusion of different covariates in the analyses.[18] To our knowledge, no published studies have examined the association between serum 25(OH)D concentration and mental health among Aboriginal and Torres Strait Islander Peoples in Australia.

The 2012-2013 Australian Aboriginal and Torres Strait Islander Health Survey (AATSIHS) assessed mental health using the Kessler Psychological Distress Scale 5 (K5) and measured serum 25(OH)D concentration using an internationally certified liquid chromatography with tandem mass spectrometry (LC-MS/MS) assay.[19, 20] Given the potential role of serum 25(OH)D concentration in mental health, we aimed to conduct exploratory research using nationally representative data collected in the AATSIHS to investigate the association between serum 25(OH)D concentration and psychological distress among Aboriginal and Torres Strait Islander Peoples in Australia.

## Methods

### Project governance

Aboriginal and Torres Strait Islander Elders and researchers are valuable knowledge holders who provide critical cultural context and guidance to research. We collaborated with Aboriginal and Torres Strait Islander Elders and researchers throughout our project to ensure that our reporting was respectful and reflective of their perspectives. Ethics approval for this study was granted by the Western Australian Aboriginal Health Ethics Committee (HREC979). The interview components of the AATSIHS were conducted under the Census and Statistics Act 1905. The biomedical component of the AATSIHS was collected under the Privacy Act 1988. Ethics approval for the biomedical component of the AATSIHS was granted at the national level by the Australian Government Department of Health and Ageing's Departmental Ethics Committee. Ethics approval for the biomedical component of the AATSIHS was granted at the jurisdictional level for New South Wales, South Australia, Western Australia, Northern Territory, and Queensland Health Service Districts.[19] Participants provided informed written consent.

**Study population**

We used cross-sectional data from the 2012-2013 AATSIHS conducted from April 2012 – July 2013, which included data from the 2012-2013 National Aboriginal and Torres Strait Islander Health Survey (NATSIHS) and biomedical data from NATSIHS participants collected as part of the 2012-2013 National Aboriginal and Torres Strait Islander Health Measures Survey.[21] The AATSIHS was conducted across Australia and included Aboriginal and Torres Strait Islander Peoples aged ≥ 2 years living in non-remote and remote areas. Of the 6,701 households approached for the NATSIHS, 5,371 (80.2%) households with a total of 9,317 participants, provided adequate responses. Comprehensive information was collected, including demographics, socioeconomic status, and mental health and wellbeing.[19]

Participants were interviewed by trained interviewers from the Australian Bureau of Statistics (ABS), and data were recorded electronically using a Computer Assisted Interview instrument.[19] In non-remote areas, up to two adults (aged ≥ 18 years) and two children per household were selected to participate in the interview. In remote areas, one adult and one child per household were selected to participate in the interview. Children aged 15 – 17 years were interviewed in person if a parent or guardian granted permission; if permission was not granted, an adult would answer questions on their behalf.

Participants aged ≥ 18 years who participated in the NATSIHS were invited to provide a blood sample for measurement of biomarkers. Blood samples were mostly collected at Sonic Healthcare clinics, from home visits, or temporary clinics at the Aboriginal Medical Services. Other pathology services, such as the Institute of Medical and Veterinary Science Pathology, were also used to collect blood samples for regional areas in South Australia and Northern Territory.

**Covariates**

The ABS reported age as a continuous variable for people aged 0 - 64 years. Due to the small number of participants in older age groups, the ABS reported only categorical data for those aged > 64 years, as follows: 65 – 69 years, 70 – 74 years, and ≥ 75 years. We assigned 67 years for all participants aged 65 – 69 years, 72 years for all participants aged 70 – 74 years, and 75 years for all participants aged ≥ 75 years.[22]

The ABS classified smokers into five categories: current smokers who smoke daily, current smokers who smoke weekly, current smokers who smoke less than weekly, ex-smokers, and never smoked (an individual who never smoked or smoked < 100 cigarettes or < 20 pipes, cigars, or other tobacco products). We re-categorised participants into two groups: (i) ex/non-smokers and (ii) current smokers. The average daily alcohol intake was calculated by ABS based on the reported number of standard drinks (10 g or 12.5 ml of alcohol) consumed over the last three drinking days. The ABS reported alcohol intake by standard drinks as a continuous variable.

We regrouped education from the 10 categories reported by the ABS into three categories: (i) no/primary/high school, (ii) diploma/certificate, and (iii) university. The ABS reported socioeconomic status in deciles according to the 2011 Index of Relative Socioeconomic Disadvantage.[19] We regrouped deciles into quintiles. A low quintile indicated a greater overall disadvantage, and a high quintile indicated a lesser overall disadvantage. The ABS classified the location of residence as non-remote (major cities, inner regional, and outer regional areas) or remote (remote and very remote areas), according to the Australian Statistical Geography Standard Remoteness Areas.[19]

It was not possible to translate the month of blood collection into the traditional calendars of the Aboriginal and Torres Strait Islander Peoples. Hence, as a proxy measure of the ultraviolet-B radiation across the year, we categorised the month of blood collection using the Western calendar definitions of Australian seasons: spring (September–November), summer (December–February), autumn (March-May), and winter (June–August).[23]

**Psychological distress**

Participants aged $\geq$ 18 years were interviewed using the K5 to assess their levels of negative emotional states over the preceding four weeks.[19] The K5 is a 5-item questionnaire, condensed and modified from the Kessler Psychological Distress Scale 10. The K5 was developed in consultation with Aboriginal and Torres Strait Islander Peoples and endorsed by the original developer of the scale.[19, 24, 25] Each item in the questionnaire is based on a five-level response scale (5: all of the time, 4: most of the time, 3: some of the time, 2: a little of the time, 1: none of the time). The result is a sum of the scores from all five questions, with a minimum score of 5 and a maximum score of 25. A score of 5 to 11 indicates low/moderate levels of psychological distress, and a score of 12 to 25 indicates high/very high levels of psychological distress.

**Measurement of serum 25(OH)D concentration**

Blood samples were sent to the Douglass Hanly Moir pathology laboratory (Sydney, New South Wales) for measurement of serum 25(OH)D concentration using an LC-MS/MS assay that was certified to the reference measurement procedures developed by the National Institute of Standards and Technology, Ghent University, and Centers for Disease Control.[19,20] Due to the small number of participants with serum 25(OH)D concentrations $\leq$ 15 and $\geq$ 130 nmol/L, the ABS reported those values as 15 and 130 nmol/L, respectively. To account for concentrations higher or lower than 15 and 130 nmol/L, we assigned all values of 15 nmol/L as 7.5 nmol/L and 130 nmol/L as 165 nmol/L.[5]

**Statistical analysis**

All statistical analyses were conducted in a secure research environment provided by the ABS.[26] Statistical analyses were performed using Stata Statistical Software: Version 18 (StataCorp, College Station, Texas).[27] We used data from participants aged $\geq$ 18 years who provided a blood sample to measure serum 25(OH)D concentration and completed the K5 questionnaire. We analysed the data using the survey weight supplied by the ABS, which weighted the data to the Aboriginal and Torres Strait Islander estimated resident population living in private dwellings of Australia at 30 June 2011, based on the 2011 Census of Population and Housing.[19]

We generated a directed acyclic graph using the 'R' package dagitty[28] to guide the selection of covariates **(Figure 1)**, resulting in the following selected covariates: age, sex, education, remoteness, socioeconomic status, season of blood collection, alcohol intake, and smoking. As body mass index may have bidirectional associations with both serum 25(OH)D concentration[29, 30] and psychological distress,[31-34] it was not adjusted for when testing the association between serum 25(OH)D concentration and psychological distress. The continuous covariates (age and alcohol intake) were tested visually for normality by plotting a frequency graph. Descriptive statistics were used to summarise participant characteristics: continuous variables were reported as mean (SD) for parametric data and median ($25^{th}$, $75^{th}$ percentile) for non-parametric data, and categorical variables were reported as frequency and percentages, n (%).

We visually evaluated the linearities between the exposure (serum 25(OH)D concentration), the outcome variable (K5), and the continuous covariates. We assessed the multicollinearity of the exposure, the outcome variable, and all covariates using the variance inflation factor; significant multicollinearity was indicated if the variance inflation factor was > 10.[35] Multivariate outlier analysis was conducted between the exposure and the outcome variable using Cook's distance to identify any influential observations.[36] Three models were generated using multiple linear regression to examine the influence of covariates on the outcome variable. Only participants with complete data for exposure, outcome, and all covariates were included in our models. Model one was unadjusted; model two was adjusted for age, sex, education, remoteness, socioeconomic status, season of blood collection, and alcohol intake; model 3 was additionally adjusted for smoking.

Given that the prevalence of vitamin D deficiency varies by remoteness [4] and the proportion of high/very high levels of psychological distress differed between males and females,[37] we further examined sex and remoteness as potential effect modifiers. We tested their interactions with serum 25(OH)D concentration and conducted stratified analyses by sex and remoteness separately across all three models. The fit of each model was assessed using R-squared, which provides the proportion of variance in each risk factor. Statistical significance was defined as p-value < 0.05.

**Results**

Of the 9,317 participants of the NATSIHS, 2,060 participated in the biomedical component. A total of 1,983 participants provided data for exposure, outcome and all covariates; of which there were 799 males and 1,184 females, 915 living in non-remote and 1,068 living in remote areas. Characteristics of participants are presented in **Table 1**.

There was no statistically significant association between serum 25(OH)D concentration and K5 in the unadjusted model **(Table 2)**. In model two, a higher 25(OH)D concentration (per 10 nmol/L) was statistically significantly associated with a lower K5 score. However, the association was attenuated after adjusting for smoking.

We did not find any interactions between serum 25(OH)D concentration and sex, or between serum 25(OH)D concentration and remoteness. The results are presented for transparent reporting and to highlight the potential exploratory value of findings with significant associations. In our sex stratified analyses, we found a significant inverse association between serum 25(OH)D concentration and K5 among females in model 2, but the association was attenuated in the final model **(Table 2)**. There were no statistically significant associations between serum 25(OH)D concentration and K5 in males across all three models. When stratified by remoteness, we found a statistically significant association between serum 25(OH)D concentration and K5 among those living in remote areas in all three models: a 10 nmol/L increase in serum 25(OH)D concentration was associated with a lower K5 score. There were no statistically significant associations between serum 25(OH)D concentration and K5 for those living in non-remote areas in all three models.

**Discussion**

Using nationally representative data from the Australian Aboriginal and Torres Strait Islander population, we found no statistically significant association between serum 25(OH)D concentration and K5 after adjusting for all covariates. However, there was an inverse association between serum 25(OH)D concentration and K5 before adjusting for smoking, suggesting that smoking may influence the observed relationship between serum 25(OH)D concentration and K5. The postulated mechanisms linking smoking and low serum 25(OH)D concentration include reduced parathyroid hormone and calcitriol production, decreased cutaneous production of vitamin D; however, the exact mechanisms are unclear.[38] Smoking may have a bidirectional association with psychological distress, as people with psychological distress may be more likely to smoke, or smokers may have higher levels of psychological distress.[39-41] Our findings are similar to other population-based studies conducted in China (n = 3,262, aged 50-70 years),[17] the United States (n = 3,916, aged ≥ 20 years),[16] and Denmark (n = 5,308, aged 18-64 years),[18] where there was no statistically significant association between circulating 25(OH)D concentration and mental health when adjusting for covariates, including smoking. Conversely, population-based studies conducted in Finland (n=5,371, aged 30-79 years)[14] and the Netherlands (n=1,282, aged 65-95 years)[13] showed an inverse association between serum 25(OH)D concentration and depression after adjusting for covariates, including smoking. Hence, the relationship between circulating 25(OH)D concentration and mental health appears complex and may be confounded by smoking in some populations.

When stratified by remoteness, we found that higher serum 25(OH)D concentration was associated with lower K5 scores only among those living remotely. There is a higher prevalence of vitamin D deficiency among Aboriginal and Torres Strait Islander Peoples living in remote (39%) compared to non-remote areas (23%), and it has been postulated that changes to clothing and housing structure since colonisation may limit sun exposure in remote areas.[42] Further, unfavourable environmental factors (e.g., accessibility to parks and lack of sporting facilities) in remote areas may limit physical activity,[43,44] resulting in reduced opportunity for sun exposure. While vitamin D can be obtained from food and supplements, dietary vitamin D intake did not differ between remote and non-remote areas in our previous analysis of the 2012-2013 National Aboriginal and Torres Strait Islander Nutrition and Physical Activity Survey,[45] and supplement use was low across the Aboriginal and Torres Strait Islander population.[46] In a previous analysis of the AATSIHS, the proportion of adults reporting high/very high levels of psychological distress was lower in those living in remote areas (24%) compared to non-remote areas (32%).[37] Aboriginal and Torres Strait Islander Peoples living in remote areas may have strong supportive community network for individuals to share their hardships, which may have beneficial impacts on their mental health.[47] It is not clear why higher serum 25(OH)D concentration was associated with lower K5 scores among those living only in remote regions; however, the higher prevalence of vitamin D deficiency and the observed association among serum 25(OH)D concentration and psychological distress among Aboriginal and Torres Strait Islander Peoples living remotely suggests that public health strategies to improve vitamin D status among this population group are needed.

Strengths of our study include the use of nationally representative data for the Aboriginal and Torres Strait Islander population and serum 25(OH)D concentration measurement using an LC-MS/MS assay that was certified to international reference measurement procedures.[19, 20] A limitation of our study was that data were collected in 2012-2013; however, the 2012-2013 AATSIHS is the most recent available nationally representative data for this population group. Given the sensitive nature of the interview, there was the potential for underreporting if participants were interviewed in the presence of other household members; Aboriginal and Torres Strait Islander Peoples living remotely may also be less inclined to provide an accurate response to an individual outside their trusted community.[19] Our study used K5, which is a non-specific measure of psychological distress, a factor that is present in a range of mental health conditions such as depression, but K5 is not a diagnostic tool.[25, 48] While useful, the K5 is limited in its ability to assess the holistic mental health and wellbeing of this population group. For Aboriginal and Torres Strait Islander Peoples, mental health and wellbeing are a combination of various factors such as their cultural identity, relationships with others, and community.[3] Future research could consider using more culturally appropriate diagnostic tools to assess individual mental health conditions and explore their association with serum 25(OH)D concentration.[49]

While we found no statistically significant association between serum 25(OH)D concentration and psychological distress among Aboriginal and Torres Strait Islander Peoples after adjusting for smoking, higher serum 25(OH)D concentration was associated with lower K5 scores among those living remotely. Adequate vitamin D status should be prioritised among Aboriginal and Torres Strait Islander Peoples, since vitamin D deficiency is prevalent among this population group.

**Data Availability Statement**

Data described in the manuscript were from national datasets found here: Australian Aboriginal and Torres Strait Islander Health Survey, detailed conditions and other health data, 2012-13, https://www.abs.gov.au/statistics/microdata-tablebuilder/datalab

**Funding**

This study was supported by the National Health and Medical Research Council (GNT1184788). BN is supported by a Curtin Strategic Scholarship.

**Ethical Approval**

Ethics approval for this study was granted by the Western Australian Aboriginal Health Ethics Committee (HREC979). The Western Australian Aboriginal Health Ethics Committee requires that all research adheres to the Australian Institute of Aboriginal and Torres Strait Islander Studies (AIATSIS) Code of Ethics for Aboriginal and Torres Strait Islander Research, which is informed by international human rights frameworks, including the Declaration of Helsinki. We used data from the 2012–2013 Australian Aboriginal and Torres Strait Islander Health Survey. The interview components of the survey were conducted under the authority of the Census and Statistics Act 1905, while the biomedical components were collected under the Privacy Act 1988.

**Table 1.** Characteristics of Aboriginal and Torres Strait Islander Peoples included in the present study (n=1,983)[1]

| Characteristics | Results |
| --- | --- |
| Age (years), median (25th, 75th percentile) | 36 (26, 48) |
| Sex, n (%) | |
|   Male | 799 (48.6) |
|   Female | 1184 (51.4) |
| Serum 25(OH)D concentration (nmol/L), mean (SD) | 64.1 (22.8) |
| K5 score, median (25th, 75th percentile) | 9 (6, 13) |
| Smoking status, n (%) | |
|   Ex/non-smoker | 1117 (64.2) |
|   Current smoker | 866 (35.8) |
| Alcohol intake, median (25th, 75th percentile)[2] | 0.5 (0, 5.8) |
| Education, n (%) | |
|   No/primary/high school | 1164 (51.3) |
|   Diploma/certificate | 681 (42.0) |
|   University | 138 (6.7) |
| Socioeconomic status, n (%) | |
|   Quintile 1 | 1187 (48.5) |
|   Quintile 2 | 348 (23.0) |
|   Quintile 3 | 200 (12.2) |
|   Quintile 4 | 181 (12.6) |
|   Quintile 5 | 67 (3.7) |
| Remoteness, n (%) | |
|   Non-remote | 915 (77.8) |
|   Remote | 1068 (22.2) |
| Season of blood collection, n (%) | |
|   Summer | 392 (27.8) |
|   Autumn | 110 (9.4) |
|   Winter | 529 (24.5) |
|   Spring | 952 (38.3) |

25(OH)D, 25-hydroxyvitamin D; K5, Kessler Psychological Distress Scale 5; SD, standard deviation.
[1]Weighted to the Aboriginal and Torres Strait Islander estimated resident population living in private dwellings of Australia at 30 June 2011, based on the 2011 Census of Population and Housing.
[2]Average daily intake based on standard drinks consumed over the last three drinking days.

**Table 2.** Association between serum 25(OH)D concentration (per 10 nmol/L) and K5 among Aboriginal and Torres Strait Islander Peoples aged ≥ 18 years[1,2]

| | Model 1 | | | | Model 2 | | | | Model 3 | | | |
|---|---|---|---|---|---|---|---|---|---|---|---|---|
| | β | 95% CI | P-value | $R^2$ | Adj β | 95% CI | P-value | $R^2$ | Adj β | 95% CI | P-value | $R^2$ |
| Total population (n=1,983) | -0.11 | -0.25, 0.02 | 0.100 | 0.004 | -0.14 | -0.28, -0.01 | 0.042 | 0.047 | -0.12 | -0.26, 0.01 | 0.074 | 0.083 |
| Stratified by sex: | | | | | | | | | | | | |
| Male (n=799) | -0.10 | -0.31, 0.11 | 0.359 | 0.003 | -0.09 | -0.29, 0.12 | 0.397 | 0.055 | -0.05 | -0.24, 0.14 | 0.590 | 0.103 |
| Female (n=1,184) | -0.12 | -0.28, 0.05 | 0.160 | 0.004 | -0.19 | -0.36, -0.01 | 0.038 | 0.043 | -0.18 | -0.36, -0.0001 | 0.050 | 0.072 |
| Stratified by remoteness: | | | | | | | | | | | | |
| Non-remote (n=915) | -0.13 | -0.28, 0.03 | 0.107 | 0.006 | -0.14 | -0.30, 0.02 | 0.082 | 0.051 | -0.12 | -0.27, 0.04 | 0.133 | 0.098 |
| Remote (n=1,068) | -0.18 | -0.37, -0.003 | 0.047 | 0.006 | -0.20 | -0.37, -0.02 | 0.025 | 0.079 | -0.18 | -0.35, -0.01 | 0.034 | 0.099 |

25(OH)D, 25-hydroxyvitamin D; β, beta coefficient; Adj β; Adjusted beta coefficient; CI, confidence interval; K5, Kessler Psychological Distress Scale 5

[1]Weighted to the Aboriginal and Torres Strait Islander estimated resident population living in private dwellings of Australia at 30 June 2011, based on the 2011 Census of Population and Housing

[2]Model 1 was unadjusted; model 2 was adjusted for age, sex, education, remoteness, socioeconomic status, season of blood collection and alcohol intake; model 3 was additionally adjusted for smoking

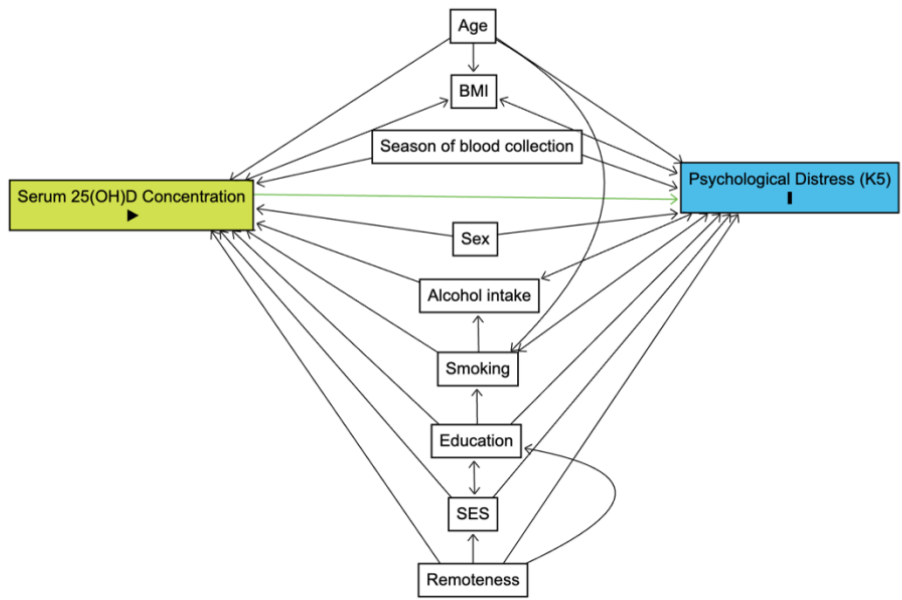

**Figure 1.** A directed acyclic graph depicting the association between serum 25(OH)D concentration and psychological distress and covariates. Green line indicates the potential associative path, solid black lines indicate a possible association, and arrows indicate the likely direction of the association.
25(OH)D, 25-hydroxyvitamin D; BMI, body mass index; SES, socioeconomic status.